\documentclass[preprint2]{aastex}
\newcommand{\myemail}{amorin@iaa.es}
\newcommand{\Ha}{\mbox{H$\alpha$}}

\newcommand{\OI}{\mbox{[\ion{O}{1}]$\lambda 6300$}\,\AA}
\newcommand{\NII}{\mbox{[\ion{N}{2}]$\lambda\lambda 6548, 6584$}\,\AA}
\newcommand{\SII}{\mbox{[\ion{S}{2}]$\lambda\lambda 6717, 6731$}\,\AA}
\newcommand{\oi}{\mbox{[\ion{O}{1}]}}
\newcommand{\nii}{\mbox{[\ion{N}{2}]}}
\newcommand{\sii}{\mbox{[\ion{S}{2}]}}
\newcommand{\ha}{\relax \ifmmode {\mbox H}\alpha\else H$\alpha$\fi}
\slugcomment{To appear in {\it The Astrophysical Journal}}
\shorttitle{Ionized gas kinematics in starbursting dwarfs}
\shortauthors{Amor\'in et al.}

\begin{document}


\title{Complex gas kinematics in compact, rapidly assembling star-forming galaxies\footnote{Based on observations made with the William Heschel Telescope operated on the island of La Palma by the Isaac Newton Group in the Spanish Observatorio del Roque de los Muchachos of the Instituto de Astrofísica de Canarias.}}

\author{R. Amor\'in \altaffilmark{1}, J.M. V\'{\i}lchez
  \altaffilmark{1}, G. F. H\"agele \altaffilmark{2,3}, V. Firpo \altaffilmark{2,3},
  E. P\'erez-Montero\altaffilmark{1}, and P. Papaderos\altaffilmark{4}
  }
%\author{R. Amor\'in \altaffilmark{1}, E. P\'erez-Montero and J.M. V\'ilchez}
\affil{(1) Instituto de Astrof\'isica de Andaluc\'ia-CSIC, Glorieta de la Astronom\'ia S/N, E-18008 Granada, Spain}
\affil{(2) Facultad de Ciencias Astron\'omicas y Geof\'{\i}sicas. Universidad de la Plata, Paseo del Bosque S/N, 1900 La Plata, Argentina}
\affil{(3) Instituto de Astrof\'isica de La Plata-CONICET, Paseo del Bosque S/N, 1900 La Plata, Argentina}
%\author{P. Papaderos\altaffilmark{2}}
\affil{(4) Centro de Astrof\'isica and Faculdade de Ci\^encias, Universidade 
do Porto, Rua das Estrelas, 4150-762 Porto, Portugal}

%\author{G. F. H\¨agele\altaffilmark{3,4}, V. Firpo\altaffilmark{3}, G. Bosch\altaffilmark{3}}
%\affil{Facultad de Ciencias Astron\´omicas y Geof\´{\i}sicas. Universidad de la Plata, Paseo del Bosque S/N, Argentina}
%\affil{Instituto de Astrof\´{\i}sica de la Plata-CONICET, Paseo del Bosque S/N, Argentina}

%\email{@}


\altaffiltext{1}{CONSOLIDER-GTC fellow; \\ 
Email: \myemail}
%\altaffiltext{2}{CSIC}
%\altaffiltext{3}{CSIC}
%\altaffiltext{4}{CAUP}


\begin{abstract}
Deep, high resolution spectroscopic observations have been obtained for 
six compact, strongly star-forming galaxies at redshift $z\sim0.1-0.3$, 
most of them also known as {\it green peas}.
Remarkably, these galaxies show complex emission-line 
profiles in the spectral region including \Ha, \NII\ and \SII, 
consisting of the superposition of different kinematical components on a 
spatial extent of few kpc: a very broad line emission underlying 
more than one narrower component. 
For at least two of the observed galaxies some of these 
multiple components are resolved spatially in their 2D-spectra, whereas 
for another one a faint detached \Ha\ blob lacking stellar continuum is 
detected at the same recessional velocity $\sim$7\,kpc away from the galaxy. 
The individual narrower \Ha\ components show high intrinsic velocity 
dispersion ($\sigma$\,$\sim$\,30--80\,km\,s$^{-1}$), suggesting together with 
unsharped masking HST images that star formation proceeds in an ensemble 
of several compact and turbulent clumps, with relative velocities of up to 
$\sim$\,500\,km\,s$^{-1}$. 
The broad underlying \Ha\ components indicate in all cases large expansion 
velocities (full width zero intensity FWZI\,$\ge$\,1000 km s$^{-1}$) and 
very high luminosities (up to $\sim$\,10$^{42}$erg s$^{-1}$), probably 
showing the imprint of energetic outflows from SNe.  
These intriguing results underline the importance of {\it green peas} for 
studying the assembly of low-mass galaxies 
% [-]
% at relatively late cosmic epochs.
% [+]
near and far.

\end{abstract}



\section{INTRODUCTION}
\label{s1}

Vigorous bursts of star-formation are key stages in the evolution of 
galaxies decisively influence their observational present and future 
integrated properties.
Theoretical studies predict some balance between significant 
gas inflow and strong star formation feedback regulating the  
growth of galaxies, especially at increasing redshifts \citep[e.g.][]{Dave12}. 
Gas accretion, either supported by small interactions/mergers with 
gas-rich companions or by gravity-driven motions produced by the 
formation and evolution of star-forming clumps in dynamically young 
systems \citep[e.g.][]{Bournaud09}, can supply the metal-poor gas 
to feed the current starburst on galactic scales. 

On the other hand, the removal of enriched gas by SNe and stellar
winds in low-mass starburst galaxies, promotes substantial chemical 
evolution \citep[e.g.][]{GTT03,RecchiHensler07}, favors the 
cessation of the current starburst episode \citep[e.g.][]{Opp-Dave06}, 
and under some conditions, could lead to positive feedback 
\citep[e.g.][]{GTT05}. 
From the observational point of view, tackling the above issues 
is extraordinary challenging and requires high quality observations. 

Studying low-mass starburst galaxies in the local Universe can 
provide key insights on the mechanisms giving origin and regulating 
enhanced star formation activity, under physical conditions approaching 
those in star-forming galaxies at higher redshifts.  
This is the case of a rare subset of low-mass galaxies at redshift 
$z$\,$\sim$\,0.1--0.3, also referred to as {\it green peas} (GP) 
\citep{Cardamone09}.
These extreme emission-line galaxies are rapidly growing systems 
characterized by their compactness, low metallicity and unusually high 
specific star formation rates (sSFR$\sim$10$^{-7}$--10$^{-9}$yr), well 
in the range of those of high-redshift galaxies \citep[e.g.][]{Bauer05}. 

In many aspects the GPs are identifiable with extreme versions of 
nearby blue compact dwarfs (BCD) galaxies, probably 
representing a major episode in their assembly history.
This conclusion relies on recent results from detailed studies on 
their physical properties and chemical abundances, integrated 
star formation histories (SFH), and photometric structure 
\citep{Amorin10,Amorin12}.  
They showed that GPs are currently producing a significant fraction 
(up to 20\%) of their total stellar mass 
($M_{\star}$\,$\sim$\,10$^{8}$--10$^{10}$\,$M_{\odot}$) 
in a galaxy-wide starburst that takes place 
over a small ($\sim$\,2--3 kpc) low-surface brightness exponential 
envelope, which might be due to more evolved stars. Extended nebular 
emission excited by a strong ongoing starburst can, however, also produce 
a large exponential envelope, mimicking a stellar disk \citep{Papaderos12}. 
Interestingly, the ionized gas-phase in these galaxies show 
low oxygen and high nitrogen-to-oxygen ratios, clearly deviating 
from the median for local galaxies of the same stellar mass.

All these properties led \citet{Amorin10} to suggest hydrodynamical 
effects e.g., massive inflows and/or enriched outflows, as playing a 
key role before and during the short and extreme phase of mass growth 
where these dwarfs are seen as GPs. 

In order to 
% [-]
% provide robust observational support on these hypothesis, 
% [+]
further explore this hypothesis, 
we are conducting a comprehensive study of the ionized gas kinematics 
and chemodynamics in these low-mass starbursts. 
In this letter we present first outstanding results on the 
remarkably complex kinematics of a handful of GPs 
observed using very deep, high resolution long-slit spectroscopy. 
%%RA: We should remove the following sentence (see report)
%% Polis: I agree: the sentence is not commented out.
% A detailed study including additional observations for a 
% larger number of galaxies will be presented in a forthcoming paper 
% (Amor\'{\i}n et al., in prep.). 

%\section{DATA}
%\label{s2}

\section{SAMPLE OF GALAXIES}
\label{s2}

The complete designations of the Sloan Digital Sky Survey (SDSS) for 
the observed sample of galaxies are included 
in Table~\ref{T1}\footnote{Through this paper we assumed a standard 
cosmology with $H_{0}$$=$70, $\Omega_{\rm \Lambda}$$=$0.7, and $\Omega_m$$=$0.3}.
The sample consists of five galaxies at $z$\,$\sim$0.2--0.3 from 
\citet{Cardamone09}, and one nearby galaxy at $z=0.1$ selected  
from a larger sample of strong emission-line galaxies 
(Amorin et al., in prep.) and included in \citet{Pilyugin12}. 
All the galaxies are very compact ($r_{50}$\,$\le$1 kpc), luminous 
($M_{\rm B}$\,$\sim$\,$-20$), and metal-poor ($Z/Z_{\odot}$\,$\sim$\,$1/5$),
rapidly star-forming systems (sSFR$=$SFR/M$_{\star}$\,$\ge$\,10$^{-9}$yr$^{-1}$) 
with no spectral signs of non-thermal ionization due to AGNs 
\citep{Cardamone09}.
%\citep[][]{Cardamone09,Amorin10,Amorin12}. 

\section{OBSERVATIONS}
\label{s3}

High-resolution spectroscopy was obtained in July 2011 as part of 
a longer-term project using the Intermediate Dispersion Spectrograph 
and Imaging System (ISIS) on the 4.2-m William Herschel Telescope 
(WHT) of the Isaac Newton Group (ING) at the Roque de los Muchachos 
Observatory (La Palma, Spain).
 We used the TEK4 CCD attached to the red arm. 
The R1200R grating was used in three different set-ups, selecting 
spectral ranges and central wavelengths around the H$\alpha$ emission 
line, depending on the redshift of the target. 
For this configuration the spatial resolution of the observations 
was 0.44 arcsec pixel$^{-1}$, and the spectral dispersion and FWHM 
effective resolution measured on the sky lines were 0.24\AA\ 
pixel$^{-1}$ and 0.52\AA, respectively. 
The spectra were taken in several exposures along the parallactic
angle, with a slit width of 0.9 arcsec. Seeing conditions varied 
between 0.7 and 1.5 arcsec during the run. Total exposure times were 
about 2 hours per galaxy.

The data was fully reduced with usual procedures (bias, overscan,
flat-fielding, co-addition and cosmic ray removal) using 
{\sc iraf}\footnote{{\sc iraf}: the Image Reduction and Analysis 
Facility is distributed by NOAO, operated by AURA, Inc., under 
agreement with the NSF}. 
Wavelength calibration was done using CuNe$+$CuAr lamp
arcs with an accuracy of about 0.02\AA. 
%%Perhaps this sentence can be removed because there is not many additional details 
Full details of the instrumental set-up, data reduction and
calibrations will be published in Amorin et al. (in prep). 

\section{RESULTS}
\label{s4}
\subsection{2D-spectra}
\label{s4.1}

\begin{figure}[t]
\epsscale{1.0}
\plotone{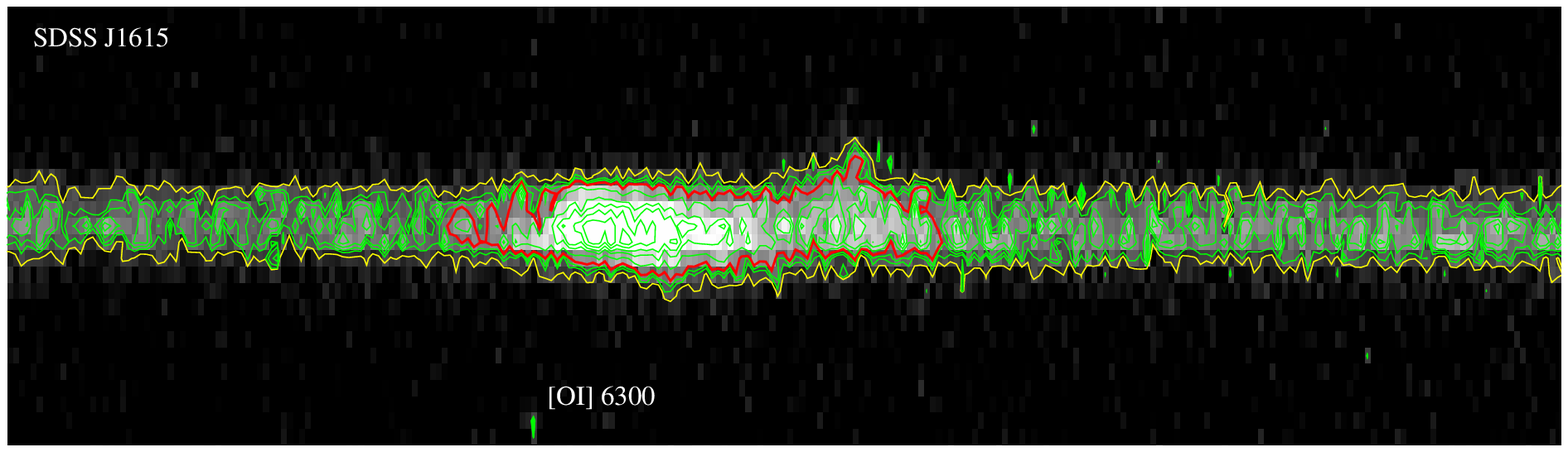}\\
\plotone{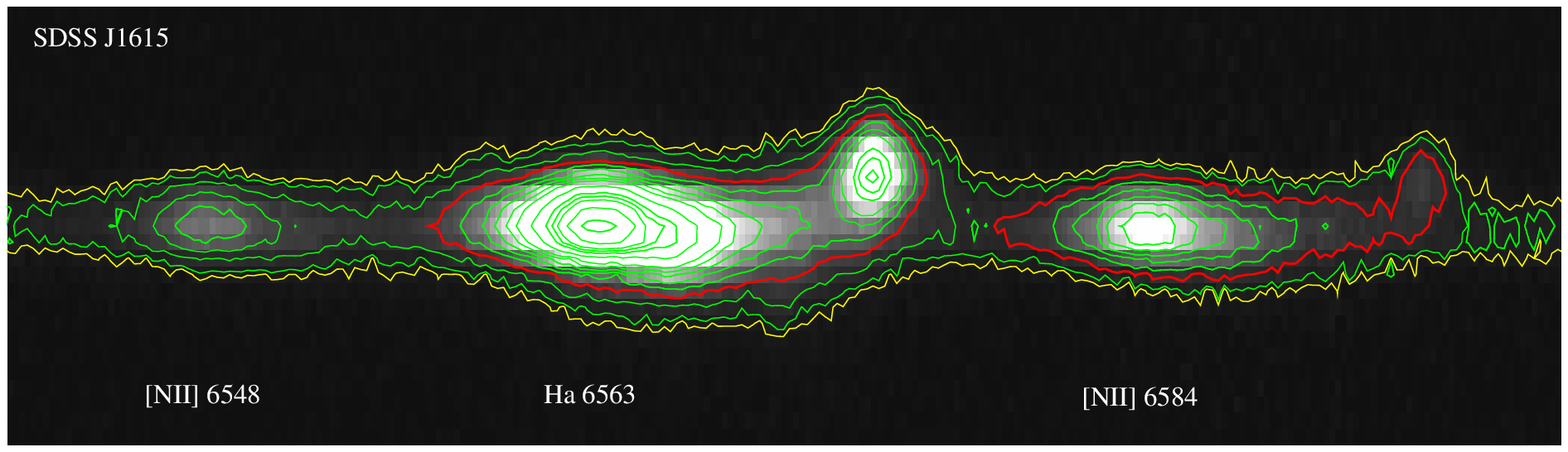}\\
\plotone{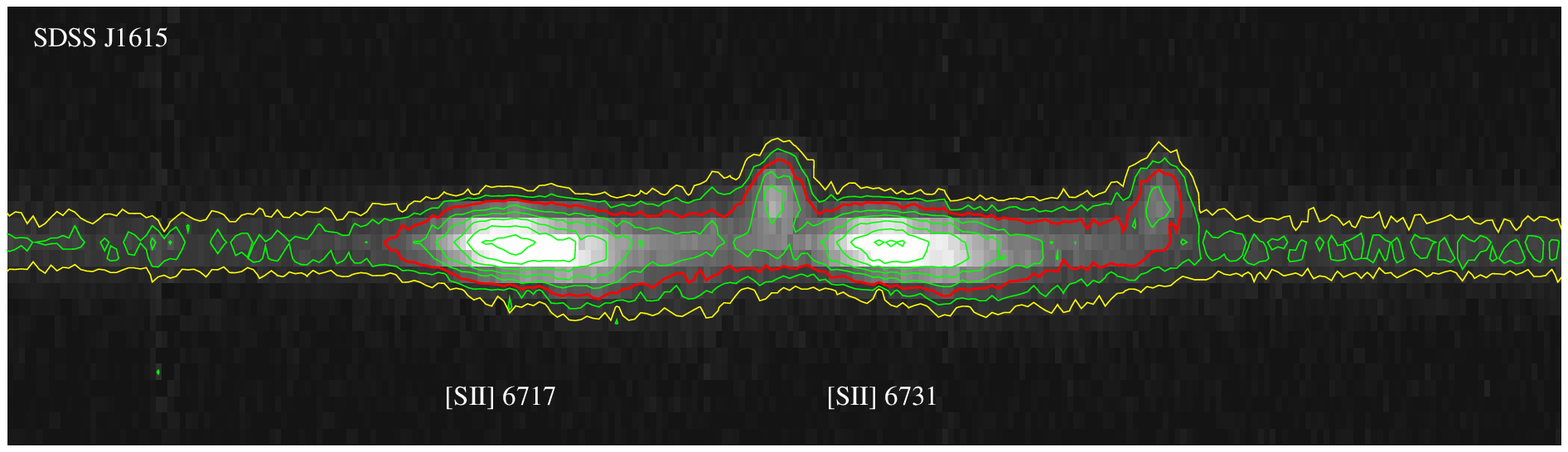}\\[.8cm]
\plotone{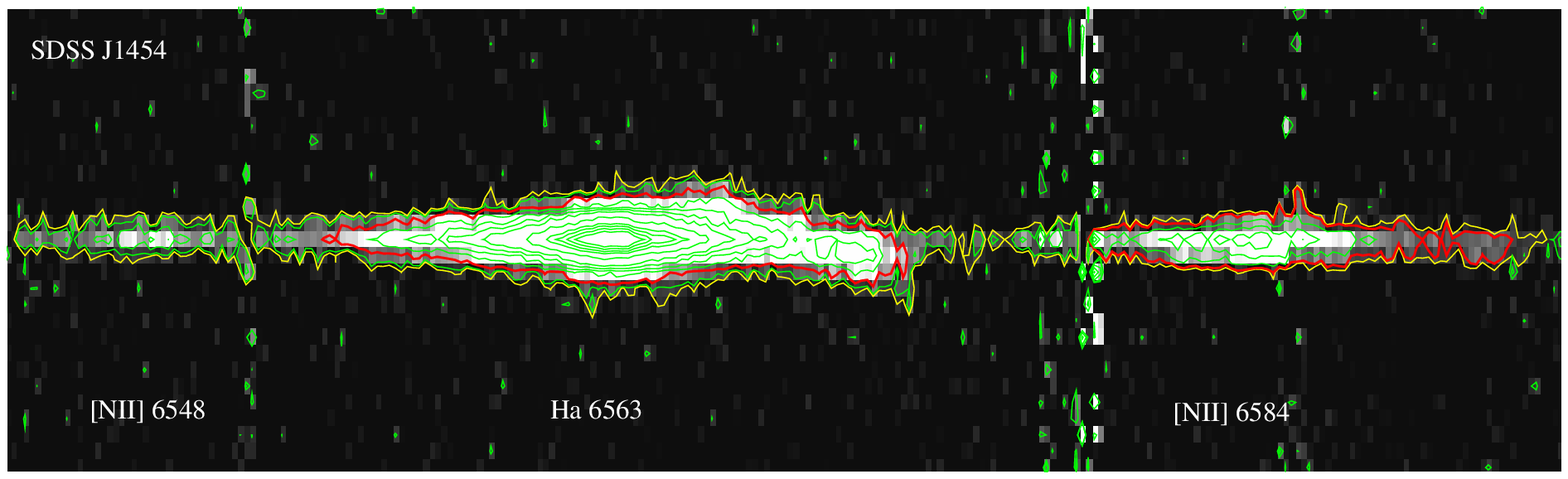}\\
\plotone{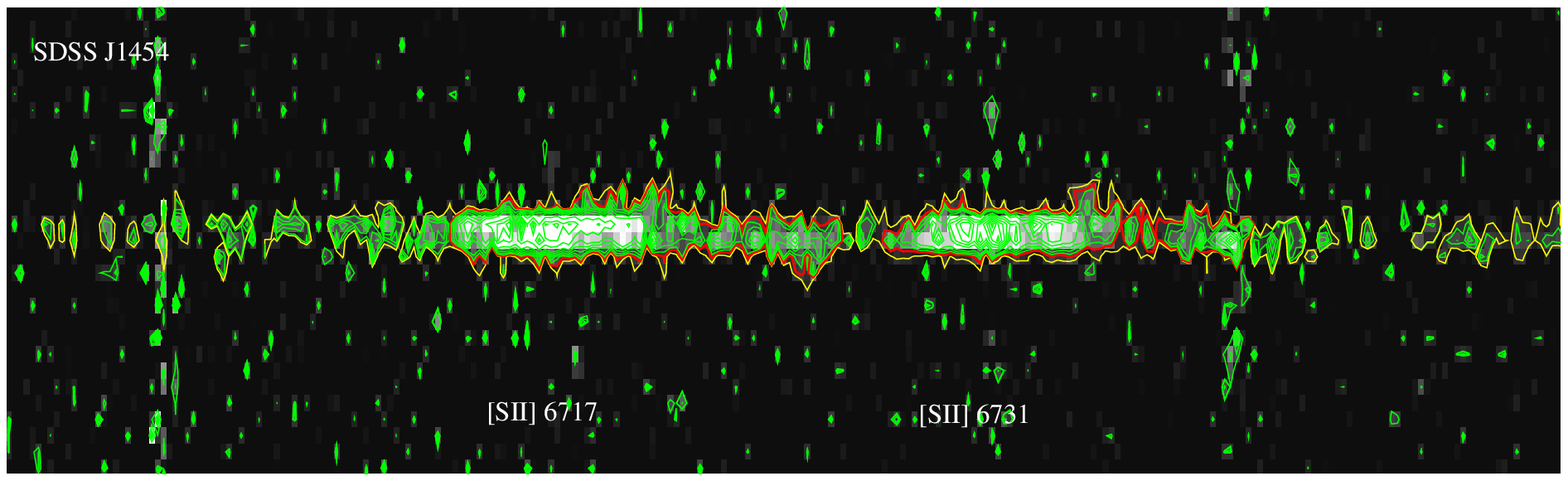}\\[.8cm]
\plotone{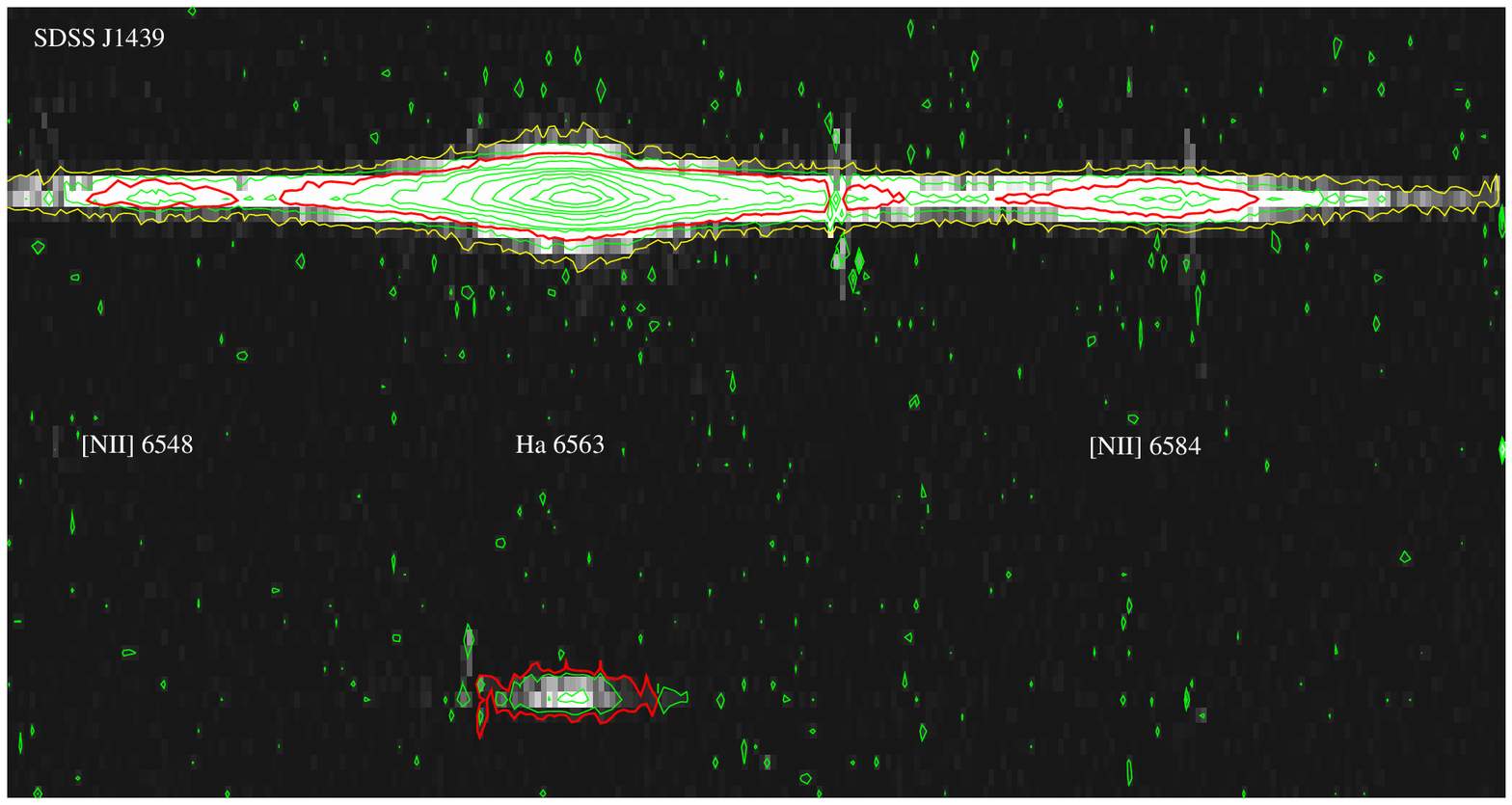}
\caption{ 2D-spectra with contours for: 
J1615 (upper panel) in the region of \oi, \ha$+$\nii, and \sii; 
J1454 in the region of \ha$+$\nii\ and \sii; and 
J1439 in the region of \ha$+$\nii.
\label{fig1}}
\end{figure}

Multiple kinematical components for each emission line can be 
identified from the 2D-spectra in all the observed galaxies. 
In Figure~\ref{fig1} we show examples of the long-slit spectrum in 
different wavelength ranges including \Ha.
In the case of J1615 we also included the \OI\ and \SII, while 
for J1454 we only included the second one since the \OI\ line is 
out of the observed spectral range.

For J1615 all the lines detected in the spectrum, even those of
lower S/N (e.g., \OI\ or \NII), show double-peaked emission. 
This feature extents to the whole optical range, as noticed by
\citet{Pilyugin12} using SDSS data. Intriguing enough, the secondary 
narrow component is offset in the spatial direction from the main 
component and the stellar continuum. 
In addition, \Ha\ in this galaxy shows very broad wings and 
a fainter component, that appears as a ``bridge'' between 
the two main narrow components. 

Though with substantially lower S/N, for J1454 
we found a similar situation to that seen in J1615. 
Two narrow components are distinguishable, at least for \Ha. 
The secondary component also appears spatially resolved. 
For the \nii\ and \sii\ doublets a hint 
of this complex spatial structure is visible as well. 

A secondary \ha\ narrow component is also spatially resolved in J1439. 
Because of its low surface brightness it is barely seen above the
broad component. 
On the other hand, an adjacent purely \ha\ emitter is projected $\sim$14 
arcsec to the SW in the slit with the same radial velocity of J1439. 
This companion at $\sim$\,7 kpc exhibits only pure \Ha\ emission in the 
spectrum, and no counterpart is detected in SDSS imaging.  
     
\subsection{Analysis of emission line profiles}
\label{s4.2}

Since it was not possible to separate the different and relatively 
close spatial components in their corresponding 1D-spectra, the 
integrated 1D-spectra in the region of \ha\ and \NII\ was used 
to analyze the structure of the emission line profiles.  
This allowed us to verify the presence of multiple components and to 
characterize their kinematics. In doing so, we adopted the technique 
presented by \citep{Hagele07} and modified later by \citet{Firpo10}. 
This technique was also successfully applied in other 
previous studies \citep[e.g.,][]{Hagele11,Hagele12,Firpo11}. 
As described in detail in \citet{Firpo10}, the method involves an 
iterative fitting of multiple Gaussian components using the task 
{\sc ngaussfit} in {\sc iraf}. Starting from the simplest solution, 
where the line profiles are formed by a broad- and a narrow-line 
component, the procedure adds extra Gaussian components until  
the solution that best fits the observed profile is obtained.
Using this technique we obtain the different Gaussian parameters 
for each component, allowing us to derive the radial velocity and 
the intrinsic velocity dispersions, the latter were corrected from 
instrumental and thermal broadening. 
We considered here an instrumental width $\sigma_i$$=$10.2 km s$^{-1}$, 
as measured from the observed lamp arcs. 
The thermal contribution was estimated assuming a kinetic temperature 
of $T=$1.2$\times$10$^4$K \citep{Amorin10,Amorin12}. 
Fluxes (and corresponding uncertainties) for each component were
derived from the amplitude and the FWHM of the Gaussian component.

\begin{figure}[ht!]
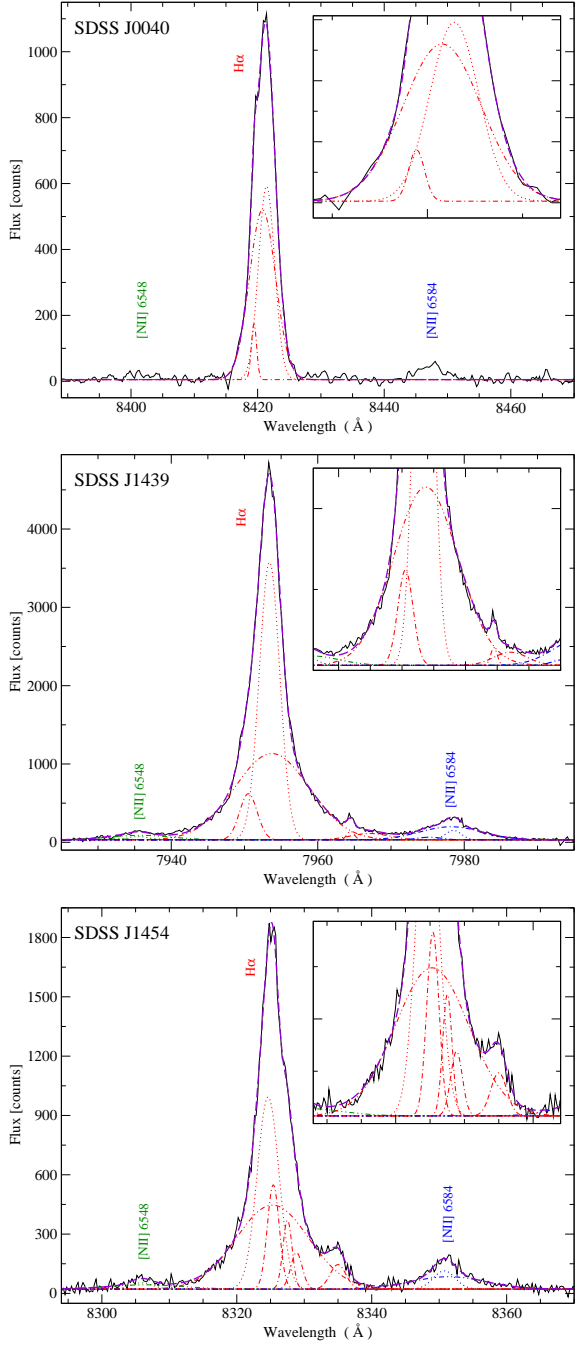

\centering
\epsscale{0.99}
\plotone{fig2_a.eps}\\[0.2cm]
\plotone{fig2_b.eps}\\[0.2cm]
\plotone{fig2_c.eps}
\caption{Multiple Gaussian fitting in the spectral range around 
\ha$+$\NII\ for the sample galaxies. The inset shows, in each case, 
a zoom of the bases of the \ha\ emission line and their  
{\sc ngaussfit} components superimposed.
\label{fig2a}}
\end{figure}

\begin{figure}[ht!]
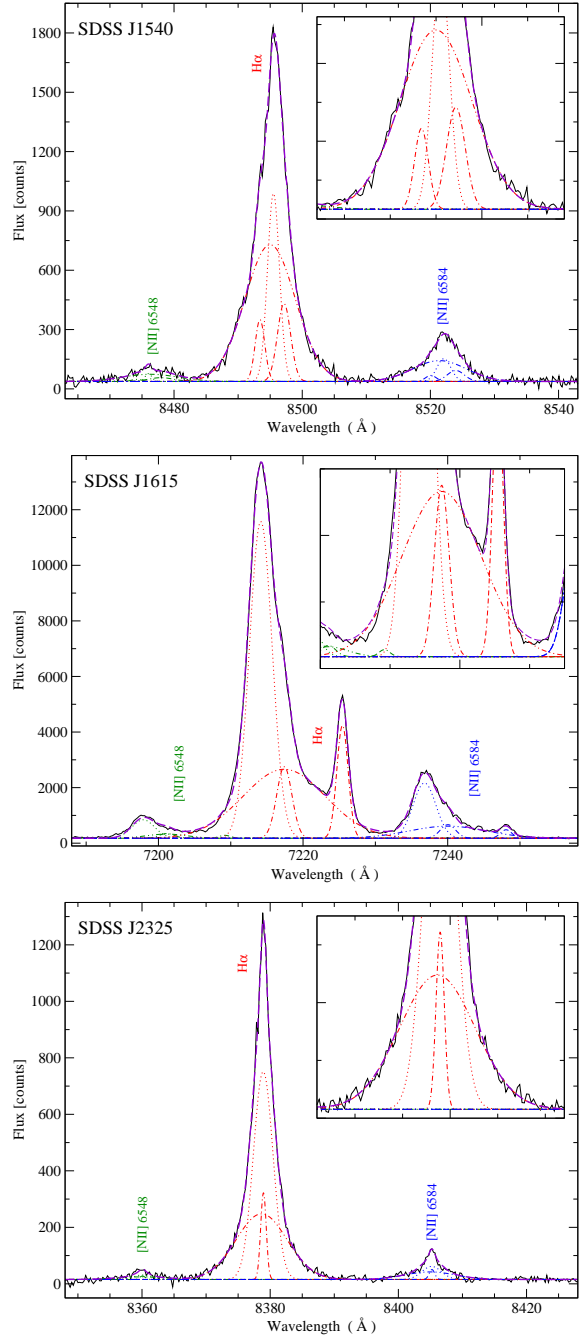

\centering
\epsscale{0.99}
\plotone{fig2_e.eps}\\[0.2cm]
\plotone{fig2_f.eps}\\[0.2cm]
\plotone{fig2_g.eps}
\caption{Same as Figure~\ref{fig2a}.
\label{fig2b}}
\end{figure}

Figures~\ref{fig2a}--\ref{fig2b} show the results of the 
{\sc ngaussfit} fitting
procedure overimposed to the observed \Ha\ and \NII\ emission lines 
for the sample. 
All the galaxies show complex line profiles, as shown by
their fit solution. 
The emission lines are well described with an underlying broad 
component and more than one narrower component. 
The kinematical centers of low-luminosity narrow and broad components 
are, in most cases, offset from those of the main narrow lines, producing 
the net effect of a clear asymmetry in the global line profiles. 
In general, and in spite of the their relatively low S/N, the 
solutions found for the \Ha\ profiles have provided a good initial guess 
for the fitting of the \nii\ doublet, whose final fits are in very 
good agreement with the \Ha\ ones. 
An exception is J0040, where \nii\ lines are too faint for a 
fitting attempt. The accuracy of the fit in the wings of the \Ha\ 
emission is highlighted in the inset of Figures~\ref{fig2a}--\ref{fig2b}.

Results from the \Ha\ fitting, including velocity dispersions and 
the fractional emission measures (in \%) for each kinematical 
component are presented in Table~\ref{T1}. 

All the six observed galaxies show a composite of more than 
one {\it strong} narrow \Ha\ components, spanning in a large 
range of velocity dispersions ($\sigma$\,$\sim$\,10--120 km s$^{-1}$) 
and luminosities (L$_{H\alpha}$\,$\sim$\,10$^{40}$--10$^{42}$ erg s$^{-1}$). 
With the exception of J2325, the \Ha\ narrow components 
are blue- or red-shifted with respect to the line centroid by 
about $\Delta$v$\sim$\,50--500 km s$^{-1}$. 
Especially remarkable are the cases of J1439, J1615, 
and J1454, which presents shifts between some of their components 
larger than 350 km s$^{-1}$. 
In two of these galaxies at least one secondary narrow
component is spatially identified in their spectra, being located on a 
spatial extent $\la$\,5--10 kpc. 

For the broad emission we found velocity dispersions and 
full width at zero intensity (FWZI) in the range of 
$\sigma$\,$\sim$\,100--250 km s$^{-1}$ and 
$\sim$\,650--1750 km s$^{-1}$, respectively. 
The corresponding broad \Ha\ luminosities are exceedingly
large $\sim$\,5$\times$10$^{41}$--1.5$\times$10$^{42}$ erg s$^{-1}$, 
representing $\sim$\,40--65\% of the total \ha\ emission (Table~\ref{T1}).
Only J1615 shows the broad emission significantly red-shifted 
($\sim$\,100 km s$^{-1}$) from the line centroid. Remarkably, forbidden 
lines, such as \nii\ also show broad components.

\section{DISCUSSION}
\label{s5}
\subsection{The broad component suggest rapid gas flows}
\label{s5.1}

The broad emission in the wings of emission lines 
suggests very high velocity gas. 
Different mechanisms have been explored in the literature to account 
for it in both giant extragalactic H{\sc ii} regions (GEHRs) 
\citep[e.g.][]{Diaz87,Castaneda90} and BCDs \citep[e.g.][]{Izotov07,James09}. 
These typically include (1) strong stellar winds caused by hot,
massive stars, e.g., WR, Ofp, and LBV stars, 
(2) expansion of multiple SNe remnants, (3) SNe-driven superbubble blow-up, 
(4) effects of turbulent mixing layers (TML), and (5) AGNs.

The presence of large amounts of WR stars has been confirmed 
for two galaxies of the sample -- J0040 and J2325 -- using 
high S/N OSIRIS-GTC spectroscopy \citep{Amorin12}. 
For some other GPs (e.g. J1615 and J1439) WR features 
are already detectable from SDSS spectra \citep{Hawley12}. 
% [+]
% Besides that, the global star formation history of the GPs \citep{Amorin12}
% {\bf and their detection in radio continuum \citep{Chakraborti12}} 
% is consistent with significant amounts of SNe. 
Besides that, the strong ongoing starburst activity in GPs \citep{Amorin12}
and their clear detection in the radio continuum \citep{Chakraborti12} 
are consistent with a significant number of WR stars and SNe. 
Both dense circumstellar envelopes of hot massive stars with strong 
stellar winds (e.g., WRs) and SNe remnants can produce broad 
components with luminosities of about 10$^{36}-$10$^{39}$ erg s$^{-1}$ 
and expansion velocities of $>1000$ km s$^{-1}$ \citep{Izotov07}. 
Our calculations confirm that the mechanical energy released
by the SNe II expected from the measured broad L(\Ha) using 
Starburst 99 models \citep{Leitherer99} at the appropriate $Z$ is 
fully consistent with their measured dispersions.
Therefore, the combined effects of (1) and (2) appear as the 
probable dominant source for the observed broad emission. 

%% in the sentence below number (3) was corrected. In the previous version was (2) which is incorrect (see report, in reference for the referee's comment number 5, third paragraph). 
Other mechanisms like (3) and (4) appear unlikely as the sole 
explanation for the broad emission at kiloparsec scales.
For example, the expansion velocity of a SNe-driven superbubble in 
blow-up phase %% (3)
is generally higher by a factor of $\sim$2-3. 
If present, TMLs do not appear to be a dominant 
effect at global scales. 
Moreover, models show TMLs only producing broad emission in Balmer 
lines but not in forbidden lines \citep{Binette09}, as observed in 
our galaxies. 
These broad components in the forbidden lines were also observed in 
circumnuclear regions \citep{Hagele07,Hagele09,Hagele10}, GEHRs 
\citep{Firpo10}, and star-forming knots of the BCD Haro~15 \citep{Firpo11}.
%% RA: Perhaps we can reword this paragraph, calling to some references. For example:
%% On the other hand, SNe-driven superbubbles may present larger uncertainties to explain the origin of the broad emission. Although they require large expansion velocities ($>$\,1000~km\,s$^{-1}$), superbubble models has some drawbacks, as explained in detail in \citep{Roy92} and \citet{Izotov07}. 
%%Similarly, TMLs do not appear to be a dominant effect at global scales. Models show TMLs only producing broad emission in Balmer lines but not in forbidden lines \citep{Binette09}, as observed in our galaxies. Broad components in the forbidden lines were also observed in circumnuclear regions \citep{Hagele07,Hagele09,Hagele10}, GEHRs \citep{Firpo10}, and star-forming knots of the BCD Haro~15 \citep{Firpo11}. Therefore (3) and (4) appear unlikely as dominant mechanisms for the broad emission observed at kiloparsec scales. 

Very broad line emission with luminosities between 10$^{40}-$10$^{42}$ 
erg s$^{-1}$ are also expected in galaxies with AGN as due to 
accretion onto an intermediate-mass black hole. 
The existence of rare low-metallicity dwarf galaxies with AGN  
have been proposed in the literature \citep[e.g.][]{Izotov07, Izotov09}.
Although we cannot rule it out, at this point the data for the GPs 
studied as yet do not support this hypothesis. 
Emission line ratios are consistent with pure starbursts and there 
is no evidence of hard non-thermal emission in none of these galaxies. 
A deeper examination of spectra looking for additional clues 
evidencing nuclear activity, e.g. presence of high ionization ions, and 
IFU data for studying the spatial distribution of the broad emission in 
these galaxies, are strongly required to reach more firm conclusions 
about this hypothesis.

Overall, the kinematics of the ISM on kpc scales in the studied 
galaxies is likely witnessing strong star formation feedback.
So we conclude that broad emission in both Balmer and forbidden lines 
are mostly originated in strong gas outflows driven by the 
intense, galaxy-wide starburst taking place in these galaxies. 

\subsection{Multiple kinematical components from emission-line fitting   
suggest multiple star-forming clumps}
\label{s5.2}

\begin{figure*}[ht]
\epsscale{0.95}
\plotone{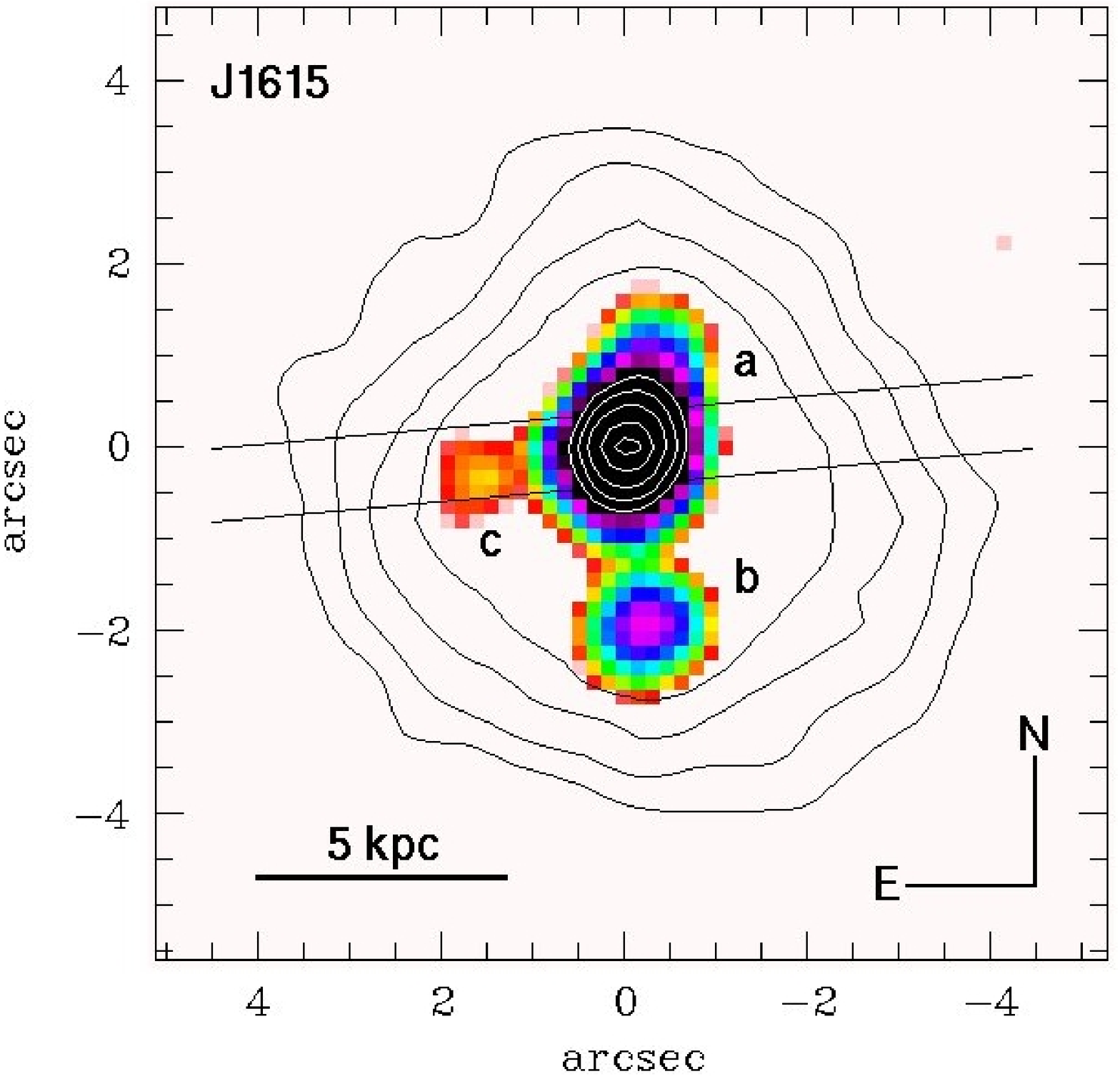}
\plotone{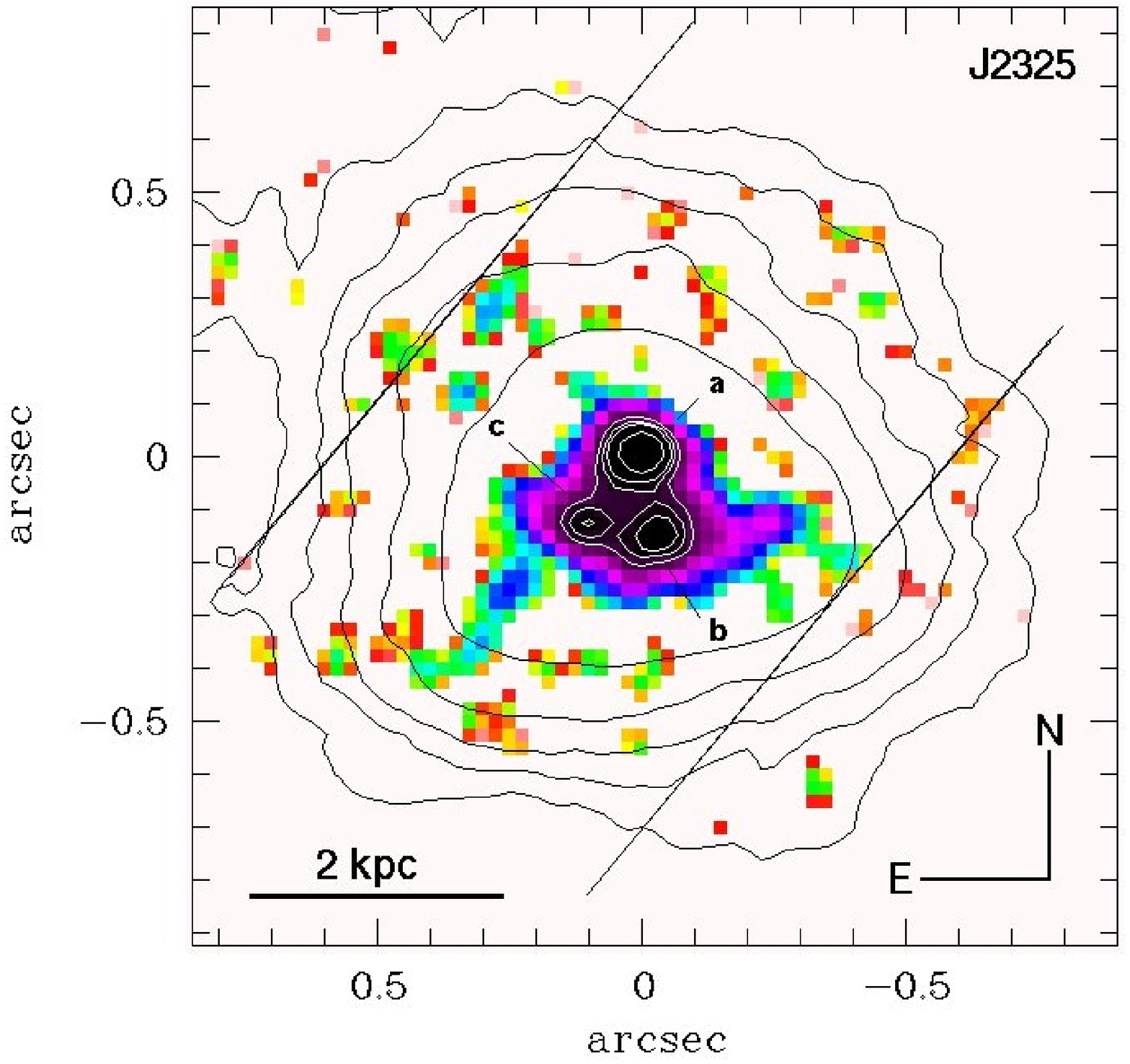}
\plotone{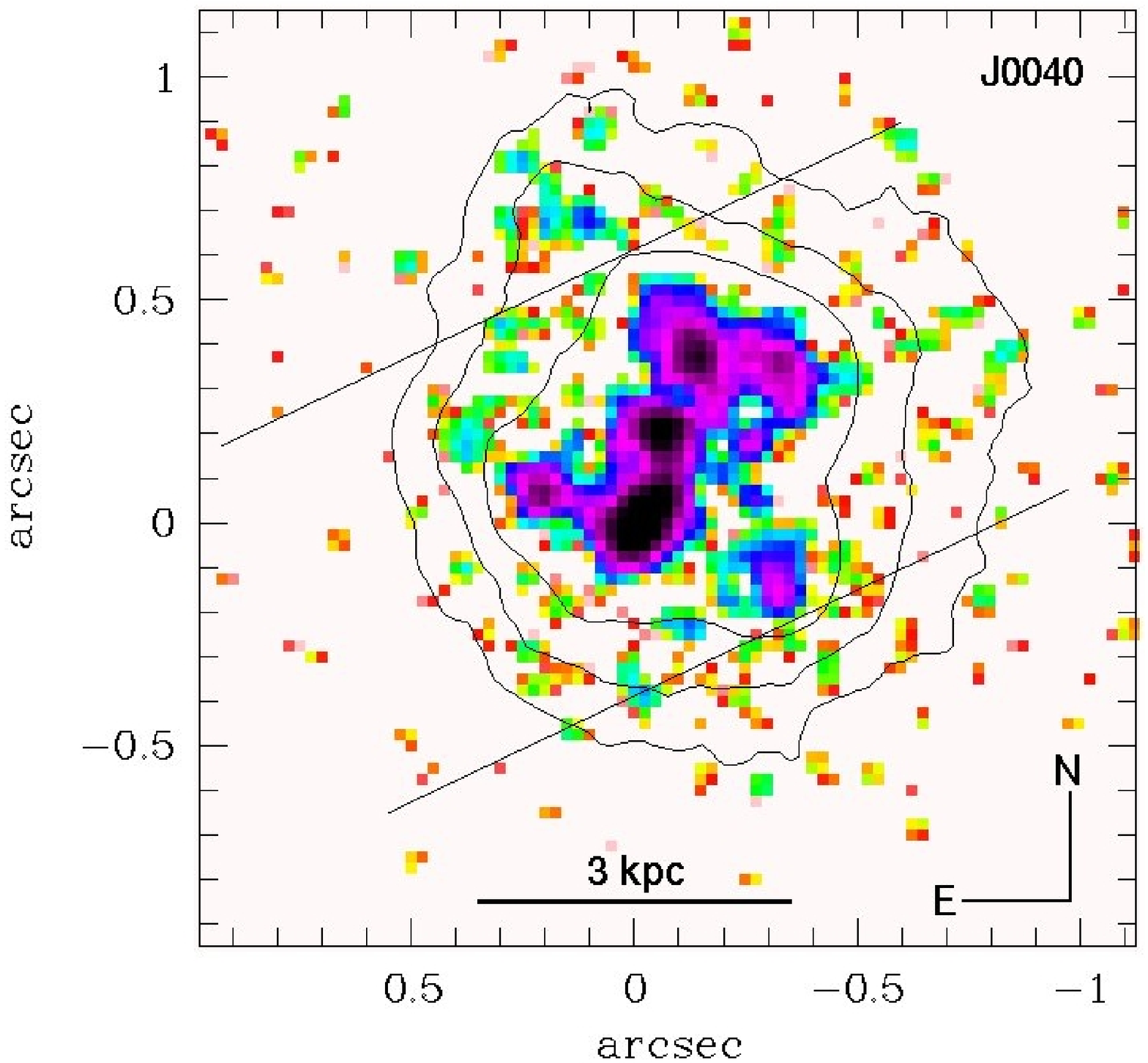}
\caption{SDSS {\it gri} (left panel) and HST/ACS F150 
(middle and right panels) of three GPs from the present sample 
after unsharp-masking.
The area covered by the longslit aperture is indicated. 
Contours are computed from the galaxy images prior to unsharp masking and 
delineate the morphology of the LSB envelope. 
North is up and east to the left. \label{fig3}}
C\end{figure*}

Decomposition of emission lines in multiple kinematical components 
with large velocity dispersions and luminosities suggest that the 
starburst episode takes place in several massive clumps. 
Ensembles of star-forming knots/clumps distributed across 
the host galaxy are typical for nearby BCDs
\citep[e.g.][]{Cairos01} 
and in compact starbursts at higher redshifts 
\citep[e.g.][]{Elmegreen05}.  
To gain further insight into the properties of the star-forming 
components in GPs, we processed archival HST/ACS 
images\footnote{HST proposal ID\,11107} of J2325 and J0040 in the 
filter F150LP with a flux-conserving unsharp-masking technique 
\citep{Papaderos98}.
The unsharp-masked images of these two GPs (Figure~\ref{fig3}) 
reveal a wealth of morphological substructure, notably three and six 
knots in J2325 and J0040, respectively, with a projected separation 
between $\sim$0.4 and $\sim$1 kpc.
Interestingly, the number of photometrically detected knots is in 
both GPs equal or larger than the kinematically distinct components 
revealed from {\sc ngaussfit}.
The compactness of GPs precludes a morphological 
analysis based on ground-based imaging from the SDSS. 
Despite this, for the nearest galaxy (J1615), the SDSS 
{\it gri} image after unsharp-masking reveails two main knots along 
the slit (a and c in Figure~\ref{fig3}), that can be associated with 
the two main narrow components in the spectrum.  

Most BCD/H{\sc ii} galaxies show emission 
lines accurately fitted using single narrow Gaussians and, eventually, 
one relatively broad component \citep[e.g.][]{Bordalo11}. 
The velocity dispersion of the {\it main narrow} component in our 
sample galaxies is considerably higher than the average 
measured turbulent velocity dispersions of GEHRs 
\citep[e.g.,][]{Firpo05,Firpo10}, 
and still higher than those in local 
BCD/H{\sc ii} galaxies of similar broad-band luminosity 
\citep[e.g.,][]{TerlevichMelnick81,Guzman96,Ostlin01,Marquart07}. 
Together with their compactness, these values are more consistent 
with those of strongly star-forming galaxies of similar 
luminosity at higher redshifts 
\citep[e.g.][]{Koo95,Wisnioski12}, rather than for nearby BCDs.

We interpret the composite profiles as an evidence of multiple massive 
star-forming clumps, distributed in a very small (few kpc) and 
dynamically young host galaxy. 
In line with results from both optical \citep{Green10} and 
near infrared integral field spectroscopy (IFS) \citep{Goncalves10} 
for UV-luminous starburst galaxies at $z$\,$\sim$\,0.1--0.3, the 
observed high velocity dispersion suggests disturbed kinematics, likely 
driven by turbulence rather than rotation. Qualitatively, 
gravity \citep[e.g.][]{TerlevichMelnick81}, 
shocks \citep[e.g.][]{GTT97}, 
accretion \citep[e.g.][]{Elmegreen10}, 
and star formation feedback \citep[e.g.][]{Lehnert09} appear as 
possible mechanisms to inject energy to drive such high velocity 
dispersion. 

\subsection{Clues for the triggering and regulation of star formation 
in ``pea'' galaxies}
\label{s5.3}

For half of the galaxies -- J1615, J1439 and J1454, --
the large difference in radial velocity found between different 
components (clumps) can possibly be interpreted as a sign of 
(clump-clump) interaction or minor-mergers.  
Mergers and tidal interactions with gas-rich, low-mass 
companions have been suggested as the main triggering mechanism 
for starbursts in nearby luminous 
BCDs \citep[e.g.,][]{Ostlin01,BergvallOstlin02}. 
These systems, closely resembling high-redshift Lyman break galaxies 
\citep{Overzier2008}, offer us a laboratory to study at high spatial resolution 
collisionally induced star formation and its r\^ole on galaxy buildup.

In the remaining galaxies, where signs of interactions are not 
obvious, gravitational instabilities may be the main cause for
the enhanced star formation. 
Models suggest that large gaseous clumps formed by gravitational 
instabilities in primordial disks can drive significant and fast
gas accretion, and trigger and sustain starburst episodes 
\citep{Bournaud07}. 
Those clumps massive enough ($\ga$\,10$^{7}$M$_{\odot}$) to resist 
disruption by star-formation can eventually coalesce toward the 
center, loss angular momentum, and form a spheroidal system 
(e.g., a bulge) in $\la$\,1 Gyr \citep{Noguchi99}.
This mechanism, has been proposed for bursty systems at 
$z$\,$\ga$\,1 \citep{Elmegreen09}, and suggested for some 
clumpy BCDs in the local Universe \citep{Elmegreen12}.

%%Perhaps we can add here a \section{SUMMARY} or \section{CONCLUSION}
In summary, observations for the six starburst galaxies presented 
here suggest that these systems are likely clumpy and highly 
turbulent, and with strong gas flows, probably as a consequence of 
strong star formation feedback. 
 
The above results highlight the important analogies found between 
the some local low-mass starbursts, like the GPs, and star-forming 
galaxies at high redshift.
A further study with additional observations and including an analysis 
of physical properties and chemical abundances for the different 
kinematical component will be presented in a forthcoming paper.  
Furthermore, high -- spectral and spatial -- resolution studies using 
IFS, tracing both star formation and gas kinematics, appear essential 
to test models and disentangle the striking gas kinematics of the GPs. 

%\section{Conclusions}
%\label{s5}

\acknowledgments
We thank the referee for his/her thoughtful comments. 
RA gratefully acknowledges the hospitality of the {\it Grupo de 
Estrellas Massivas} at the IALP/FCAGLP, where part of this work was 
carried out.    
We thank G. Bosch and A. Monreal-Ibero for helpful suggestions on 
the data analysis. 
We also thank C. Mu\~noz Tu\~n\'on, G. Tenorio-Tagle, and 
R. Terlevich for useful comments. 
This work has been funded by grants, 
AYA2010-21887-C04-01: 
{\it Estallidos de Formaci\'on Estelar en Galaxias} 
(\url{http://estallidos.iac.es/estallidos/}), 
and CSD2006-00070: {\it First Science with the GTC} 
(\url{http://www.iac.es/consolider-ingenio-gtc/}) of the
Consolider-Ingenio 2010 Program, by the Spanish MICINN.
P. Papaderos is supported by a Ciencia 2008 contract, funded 
by FCT/MCTES (Portugal) and POPH/FSE (EC).


%% in parentheses, after the keyword, but they will not be verified.

{\it Facilities:}\facility{WHT (ISIS)}, \facility{HST (ACS)}, \facility{SDSS}.

\clearpage

%\begin{figure}[ht]
%\epsscale{0.98}
%\plotone{./figures/Obj4_perfil-Ha.eps}
%\caption{ Spatial profile of the light distribution along the slit for the 
%observed \ha\ emission in SDSS J1439. 
%The profiles correspond to line$+$continuum (dashed line), continuum 
%(dashed-dotted line), and the difference between them (solid line), 
%representing the pure emission from \ha.  
%The inset shows a zoom of the lower-flux levels. 
%See the electronic edition of the Journal for a color version 
%of this figure.\label{fig2}}
%\end{figure}

\clearpage

\clearpage
\clearpage

\begin{deluxetable}{ccccc}
\tabletypesize{\scriptsize}
\tablecaption{\Ha\ broad and narrow components \label{T1}}
\tablecolumns{4} 
\tablewidth{0pt}
\tablehead{
\colhead{Galaxy\tablenotemark{\dagger}}&\colhead{$\Delta$v\tablenotemark{a}}
&\colhead{$\sigma$\tablenotemark{b}}
&\colhead{FWZI\tablenotemark{c}}
&\colhead{EM$_f$\tablenotemark{d}}\\
%&\colhead{L$_{H\alpha}$\tablenotemark{d}} \\
\colhead{component}&\colhead{km s$^{-1}$}
&\colhead{km s$^{-1}$}
&\colhead{(\AA) km s$^{-1}$}
%&\colhead{10$^{41}$ erg s$^{-1}$}
&\colhead{\%}
}
%\tableline
\startdata
\sidehead{SDSS J004054.31+153409.8 (=J0040), $z$\,$=$0.283 }
broad & -23$\pm$2 &  94$\pm$2 & (18) 640 & 56 \\
n1    &   0$\pm$1 &  57$\pm$1 & & 40 \\
n2    & -71$\pm$2 &   9$\pm$1 & & 4 \\
\sidehead{SDSS J143905.23+245353.3 (=J1439), $z$\,$=$0.212 }
broad&  12$\pm$3  & 234$\pm$2  & (39) 1470 & 49 \\
n1   &   0$\pm$1  &  62$\pm$1  & & 43 \\
n2   &-109$\pm$2  &  53$\pm$2  & & 6 \\
n3   & 418$\pm$6  &  16$\pm$5  & & 1 \\
n4   & 520$\pm$20 & 120$\pm$20 & & 2 \\
\sidehead{SDSS J145435.57+452856.3 (=J1454), $z$\,$=$0.268 }
broad&  24$\pm$3 & 248$\pm$2 & (37) 1330 & 47 \\
n1   &   0$\pm$1 &  73$\pm$1 & & 32 \\
n2   &  28$\pm$2 &  37$\pm$1 & & 10 \\
n3   & 104$\pm$2 &  26$\pm$2 & & 5 \\
n4   & 152$\pm$2 &  35$\pm$2 & & 3 \\
n5   & 373$\pm$6 &  51$\pm$3 & & 3 \\
\sidehead{SDSS J154050.19+572441.9 (=J1540), $z$\,$=$0.294 }
broad&-18$\pm$2 & 180$\pm$2 & (30) 1060 & 64 \\
n1   &  0$\pm$1 &  39$\pm$1 & & 21 \\
n2   & 58$\pm$2 &  44$\pm$2 & & 9 \\
n3   &-74$\pm$2 &  30$\pm$2 & & 6 \\
\sidehead{SDSS J161555.12+420624.5 (=J1615), $z$\,$=$0.100 }
broad& 134$\pm$4 & 264$\pm$2 & (42) 1750 & 38 \\
n1   &   0$\pm$1 &  71$\pm$1 & & 47 \\
n2   & 135$\pm$2 &  46$\pm$2 & & 7 \\
n3   & 469$\pm$2 &  32$\pm$2 & & 8 \\
\sidehead{SDSS J232539.22+004507.2 (=J2325), $z$\,$=$0.277 }
broad& -9$\pm$3  &  183$\pm$3& (31) 1110 & 42 \\
n1   &  0$\pm$1  &   69$\pm$1& & 52 \\
n2   &  2$\pm$1  &   10$\pm$1& &  6 \\
\enddata
%\tableline
\tablenotetext{\dagger}{Complete name and redshift from the SDSS.}
\tablenotetext{a}{$\Delta$v$=$v$_{obs}-$v$_{comp}$, where v$_{obs}$ and v$_{comp}$ are the velocity at the peak of the fitted H$\alpha$ emission line and the main narrow component, respectively.} 
\tablenotetext{b}{The intrinsic velocity dispersion, $\sigma$, is calculated here as $\sigma$\,$=$\,$(\sigma^2_{g} - \sigma^2_{i} - \sigma^2_{t})^{1/2}$, where $\sigma_{g}$ is the velocity dispersion from the fitted Gaussian, and $\sigma_{i}$ and $\sigma_{t}$ are the instrumental and the thermal broadening, respectively.}
\tablenotetext{c}{1\AA\ at $\lambda_{H\alpha}=$6562.8\AA$\sim$45.7$(1+z)^{-1}$km/s. }
\tablenotetext{d}{Fractional emission measures (EM$_f$) in per cent.}

\end{deluxetable}
%%%%%%%%%%%%%%%%%%%%%%%%%%%%%%%%%%%%%%%%%%%%%%%%%%%%%%%%%%%%%%%%%%%%%%%%%%%%

\end{document}